\documentclass[aps,prl,twocolumn,floatfix,showpacs]{revtex4}
\usepackage{amsmath,amssymb,graphicx,bm}
\begin{document}
\title{Absence of the diffusion pole in the Anderson insulator}

\author{V.  Jani\v{s}} \author{ J. Koloren\v{c}}

\affiliation{Institute of Physics, Academy of Sciences of the Czech
  Republic, Na Slovance 2, CZ-18221 Praha 8, Czech Republic}
\email{janis@fzu.cz, kolorenc@fzu.cz}

\date{\today}

%\maketitle

\begin{abstract}
We discuss conditions for  the existence of the diffusion pole and its
consequences in disordered noninteracting electron systems. Using only
nonperturbative and exact arguments we find against expectations that the
diffusion pole can exist only in the diffusive (metallic) regime. We
demonstrate that the diffusion pole in the Anderson localization phase
would lead to nonexistence of the self-energy and hence to a physically
inconsistent picture. The way how to consistently treat and understand the
Anderson localization transition with vanishing of the diffusion pole is
presented.
\end{abstract}

\maketitle %newpage
The Anderson insulator is a specific disordered or amorphous solid with
available quasiparticle states at the Fermi surface but with no diffusion
or charge transport at long distances. Possibility of the absence of
diffusion in impure metals and alloys was proposed by P. W. Anderson on a
simple tight-binding model of disordered noninteracting electrons
\cite{Anderson58}. Since then, vanishing of diffusion, now called Anderson
localization, has attracted much attention of both theorists and
experimentalists~\cite{Kramer93}. In spite of a considerable portion of
amassed experimental data, disclosed various specific and general aspects
of the Anderson metal-insulator transition and a~number of theoretical and
computational approaches so far developed, we have not yet reached
complete understanding of Anderson localization. Although the basic
aspects of the critical behavior at the Anderson localization transition
are known, the position of this disorder-driven metal-insulator transition
within the standard classification scheme of phase transitions with
control and order parameters remains yet unclear.

In this Letter we reexamine the description of Anderson localization with
averaged functions. Configurational averaging is an important means to
restore translational invariance in disordered systems and to provide
reproducibility (sample independence) of the findings. Disordered systems
after averaging behave as pure ones with effective correlations between
the motion of individual quasiparticles. The translationally invariant
description is hence the proper tool for comprehending the Anderson
localization transition in the way we understand phase transitions in
clean interacting systems.
 
The aim of the Letter is to set forth general constraints on the
translationally invariant description of the Anderson localization
transition resulting from exact relations, conservation laws and equations
of motion for the averaged Green functions. We explicitly demonstrate that
the weight of the diffusion pole cannot be fixed to unity as dictated by
the Ward identity and that the diffusion pole must vanish in the localized
phase. The diffusion pole in the localized phase would lead to
nonexistence of the self-energy and to an unsolvable equation of motion
for the two-particle vertex.

The fundamental building blocks of the translational invariant description
of disordered electrons are averaged one and two-particle resolvents
$G(\mathbf{k},z)$ and $G^{(2)}_{{\bf k}{\bf k}'}(z_+,z_-;\mathbf{q})$,
respectively. Here $\mathbf{k}, \mathbf{q}$ are fermionic and bosonic
momenta and $z_\pm$ are complex energies. We use the electron-hole
representation for the two-particle Green function with $\mathbf{k}$ and
$\mathbf{k}'$ for incoming and outgoing electron momenta. The bosonic
momentum $\mathbf{q}$ measures the difference between the incoming momenta
of the electron and the hole.

Fundamental functions for the description of a linear response of an
electron gas to an external electromagnetic perturbation are the density
response and the electron-hole correlation function defined from the
two-particle resolvent as
\begin{subequations}\label{eq:DRF}
\begin{align} \label{eq:DRF-thermo}
  \chi({\bf q},i\nu_m) = -\frac 1{\beta N^2}\sum_{{\bf k}{\bf k}'}
\sum_{n=-\infty}^{\infty} G^{(2)}_{{\bf k}{\bf k}'}(i\omega_n,i\omega_n +
 i\nu_m;{\bf q}), \\ \label{eq:DRF-Phi}   \Phi^{RA}_E(\mathbf{q},\omega) =
\frac 1{N^{2}}\sum_{\mathbf{k}\mathbf{k}'}
G^{(2)}_{\mathbf{k}\mathbf{k}'}(E+ \omega + i0^+ , E - i0^+;\mathbf{q}) \
.
\end{align}
\end{subequations}
For convenience, the density response $\chi$ was defined at the
temperature axis, while the electron-hole correlation function $\Phi^{RA}$
for real frequencies. The superscript indices $R,A$ relate to
the limits of complex energies from which the first and second real energy
variables were reached.

It is useful to go over from the full two-particle resolvent to a
two-particle vertex $\Gamma$ defined as
\begin{multline}\label{eq:G2-Gamma}
G^{(2)}_{{\bf k}{\bf k}'}(z_+,z_-;\mathbf{q}) = G(\mathbf{k} +
\mathbf{q},z_+)G(\mathbf{k},z_-)\left[1 \right. \\ \left. +
\Gamma_{{\bf k}{\bf k}'}(z_+,z_-;\mathbf{q}) G(\mathbf{k}' + \mathbf{q}
,z_+) G(\mathbf{k}',z_-)\right]\ .
\end{multline}
The two-particle vertex obeys a Bethe-Salpeter equation of motion. Since we
use the electron-hole representation, the most natural way to construct the
Bethe-Salpeter equation is to use ladders in the electron-hole scattering
channel. We then can write
\begin{multline}\label{eq:BS-eh}
\Gamma_{\mathbf{k}\mathbf{k}'}(\mathbf{q}) =
\Lambda^{eh}_{\mathbf{k}\mathbf{k}'}(\mathbf{q})\\ + \frac
1N\sum_{\mathbf{k}''}
\Lambda^{eh}_{\mathbf{k}\mathbf{k}''}(\mathbf{q})
G_+(\mathbf{k}'' + \mathbf{q}) G_-(\mathbf{k}'' )
\Gamma_{\mathbf{k}''\mathbf{k}'}(\mathbf{q})
\end{multline}
where we introduced the electron-hole irreducible vertex $\Lambda^{eh}$. We
suppressed the energy variables in Eq.~\eqref{eq:BS-eh}, since they are not
dynamical quantities and act only as external parameters. They are easily
deducible from the one-electron propagators $G_\pm(\mathbf{k}) \equiv
G(\mathbf{k},z_\pm)$.

We have up to now used only definitions or representations of the
two-particle functions. The first important equation demanding a proof is
the Vollhardt-W\"olfle identity expressing conservation of the norm of the
wave function of a free electron scattered on a static impurity
potential
\begin{multline}
  \label{eq:VWW-identity}   \Sigma^R({\bf k},E + \omega) - \Sigma^A({\bf
k}, E) = \frac 1N \sum_{{\bf k}'}\Lambda^{RA}_{{\bf k}{\bf k}'}(E +
\omega, E) \\ \times \left[G^R({\bf k}',E + \omega) - G^A({\bf k}', E)
\right] \ .
\end{multline}
Here we denoted $\Sigma^{R},\Sigma^{A}$ the retarded and advanced
self-energy and $\Lambda^{RA}_{{\bf k}{\bf k}'}(E + \omega, E ) \equiv
\Lambda^{eh}_{{\bf k}{\bf k}'}(E + \omega + i0^+, E- i0^+;\mathbf{0})$.
Equation \eqref{eq:VWW-identity} was proved diagrammatically in
Ref.~\cite{Vollhardt80b}. A nonperturbative  proof exists trough
equivalence of Eq.~\eqref{eq:VWW-identity} with the Velick\'y identity
\cite{Velicky69,Janis01b}. We hence can consider
Eq.~\eqref{eq:VWW-identity} as a nonperturbative property of solutions of
the Anderson model of disordered electrons \cite{Note1}.

The first fundamental consequence of Eq.~\eqref{eq:VWW-identity} is the
existence of the diffusion pole in the electron-hole correlation function
$\Phi^{RA}$. Introducing a quantum diffusion function
$D(\mathbf{q},\omega)$ we can represent the density response in the
hydrodynamic limit $q\to 0$ as \cite{Janis02a}
\begin{equation}
  \label{eq:density-diffusion}
  \chi(\mathbf{q},\omega) \doteq \frac {\chi_0 D(\omega)
    q^2}{-i\omega + D(\omega) q^2} + O(q^2)
\end{equation}
where $\chi_0 = \lim_{q\to0}\lim_{\omega\to0} \chi(\mathbf{q},\omega)$ and
$D(\omega) = D(\mathbf{0},\omega)$. The dynamical diffusion coefficient
$D(\omega)$  reduces in the static limit to the diffusion constant related
at zero temperature to the static optical conductivity by the Einstein
formula $D = D(0) = \sigma/e^2 n_F$, where $n_F$ is the density of states
at the Fermi energy \cite{Janis02a}.

Using Eqs.~\eqref{eq:DRF} we can now expand the density response in the
hydrodynamic ($q\to0$) and quasistatic ($\omega\to0$) limits maintaining
$D(\omega)q^2 \sim \omega$. We find that its leading term at zero
temperature is governed by the electron-hole correlation function
\begin{equation}
\label{eq:DRF-0} \chi(\mathbf{q},\omega) = \chi(\mathbf{q},0) +
\frac{i\omega}{2\pi} \left( \Phi^{RA}_{E}(\mathbf{q},\omega) +
O(q^0)\right)\\ + O(\omega)\ .
\end{equation}
Inserting the asymptotic expansion Eq.~\eqref{eq:DRF-0} in
Eq.~\eqref{eq:density-diffusion} we derive an explicit representation of
the diffusion pole as the leading low-energy term in the electron-hole
correlation function
\begin{equation}
  \label{eq:Phi-diffusion}
  \Phi^{RA}_{E}({\bf q},\omega)\doteq  \frac {2\pi n_F}{-i\omega +
    D(\omega)q^2} + O(q^0,\omega^0)\ .
\end{equation}
Although the denominator in the density response in
Eq.~\eqref{eq:density-diffusion} is identical with that from the
electron-hole correlation function in Eq.~\eqref{eq:Phi-diffusion}, it is
only the latter function that is singular in the low-energy limit. This
singularity, the diffusion pole, is a consequence of the particle-number
conservation expressed mathematically in the Ward identity
\eqref{eq:VWW-identity}. The only parameter depending on the character of
the underlying solution in Eq.~\eqref{eq:Phi-diffusion} is the dynamical
diffusion coefficient $D(\omega)$. This simplified low-energy asymptotics
of the electron-hole correlation function is the basis for the existing
scaling and field-theoretic approaches to the problem of Anderson
localization \cite{Lee85}.

According to the above reasoning the diffusion pole must exist also in the
state with localized electrons, where the dynamical diffusion
coefficient is expected to vanish in the static limit as
\cite{Vollhardt80b}
\begin{equation}\label{eq:diffusion-localized}
D(\omega)\sim -i\omega \xi^2
\end{equation}
where $\xi$ is the localization length. We, however, show that such a
picture is not sustainable, since the diffusion pole in the localized
phase would hinder the existence of a finite self-energy almost everywhere
within the energy band.

To manifest this we first show that in systems invariant with respect to
time reversal the diffusion pole enters the electron-hole irreducible
vertex $\Lambda^{eh}$ as the so-called Cooper pole. It displays the
same low-energy asymptotics as the electron-hole correlation function
$\Phi^{RA}$.

Time inversion in noninteracting systems with elastic scatterings only is
expressed as inversion of the particle propagation
$\mathbf{k}\to-\mathbf{k}$, i.~e., the electron and hole interchange their
roles. The time-reversal invariance then means that $G(\mathbf{k},z) =
G(-\mathbf{k},z)$. For the two-particle resolvent we then obtain either
$G^{(2)}_{\mathbf{k}\mathbf{k}'}(z_+,z_-;\mathbf{q}) =
G^{(2)}_{-\mathbf{k}' -\mathbf{k}}(z_+,z_-;\mathbf{q} + \mathbf{k} +
\mathbf{k}')$ or equivalently
$G^{(2)}_{\mathbf{k}\mathbf{k}'}(z_+,z_-;\mathbf{q}) = G^{(2)}_{\mathbf{k}
\mathbf{k}'}(z_+,z_-;-\mathbf{q} - \mathbf{k} - \mathbf{k}')$ leading to
$\Gamma_{\mathbf{k}\mathbf{k}'}(\mathbf{q}) = \Gamma_{-\mathbf{k}'
-\mathbf{k}}(\mathbf{q} + \mathbf{k} + \mathbf{k}')$. However, the
electron-hole irreducible vertex $\Lambda^{eh}$ is not invariant in the
same way the full two-particle vertex $\Gamma$ is. We actually have
$\Lambda^{eh}_{-\mathbf{k}' -\mathbf{k}}(\mathbf{q} + \mathbf{k} +
\mathbf{k}')=\Lambda^{ee}_{\mathbf{k}\mathbf{k}'} (\mathbf{q})$, where
$\Lambda^{ee}$ is the electron-electron irreducible vertex being generally
different from the electron-hole one. The Bethe-Salpeter equation
\eqref{eq:BS-eh} after time reversal is hence transformed to another
Bethe-Salpeter equation with the electron-electron irreducible vertex and
modified momentum convolutions. The existence of several Bethe-Salpeter
equations for the same full two-particle vertex $\Gamma$ is a well known
ambiguity in the definition of the two-particle irreducibility
\cite{Janis01b}.
 
The low-energy singularity in the electron-hole correlation function is a
consequence of the existence of a singularity in the full two-particle
vertex $\Gamma^{RA}$, since the former function is the latter one
multiplied with one-electron propagators and summed over fermionic momenta,
cf. Eqs.~\eqref{eq:DRF-Phi} and \eqref{eq:G2-Gamma}. It means that
$\Gamma^{RA}_{\mathbf{k}\mathbf{k}'}(\mathbf{q})$ diverges with
$\omega,\mathbf{q}^2 \to 0$ for almost all fermionic momenta
$\mathbf{k},\mathbf{k}'$ so that its singularity survives in $\Phi^{RA}$
as the diffusion pole, Eq.~\eqref{eq:Phi-diffusion}. However, due to the
electron-hole symmetry, the full two-particle vertex must show the same
singularity also in the limit $\omega,(\mathbf{k} + \mathbf{k}' +
\mathbf{q})^2 \to 0$.

Further on, the (leading) singularities in the full vertex $\Gamma$ are
already contained in the irreducible vertices $\Lambda^{eh}$ and
$\Lambda^{ee}$. This conclusion follows from the so-called parquet
equation. It is an expression of topological nonequivalence of different
two-particle irreducibilities.  That is, the reducible function of one type
(convolution of two or more irreducible vertices of the same type) is
irreducible in the other irreducibility channels. In case of two
irreducibility channels (two types of Bethe-Salpeter equations) we can
write the parquet equation as
\begin{equation}\label{eq:parquet-equation}
\Gamma_{\mathbf{k}\mathbf{k}'}(\mathbf{q}) =
\Lambda^{ee}_{\mathbf{k}\mathbf{k}'}(\mathbf{q}) +
\Lambda^{eh}_{\mathbf{k}\mathbf{k}'}(\mathbf{q}) -
\mathcal{I}_{\mathbf{k}\mathbf{k}'}(\mathbf{q}) \ .
\end{equation}
where $\mathcal{I} = \Lambda^{eh} \cap \Lambda^{ee}$ is a vertex
irreducible in both irreducibility channels (completely irreducible
vertex) \cite{Janis01b}.

We deduce the form of the pole in the electron-hole irreducible vertex
$\Lambda^{eh}$ from the asymptotic limit to high spatial dimensions
\cite{Janis04a}. There we obtain that the completely irreducible vertex
$\mathcal{I}$ is local and regular. The singularity of the full vertex
$\Gamma^{RA}$ in the limit $\omega,\mathbf{q}^2 \to 0$ is contained in the
electron-electron irreducible vertex, $\Lambda^{ee}$, while the singularity
for $\omega,(\mathbf{k} + \mathbf{k}' + \mathbf{q})^2 \to 0$ appears only
in $\Lambda^{eh}$. The low-energy singularities of the irreducible
vertices are fixed in momentum space by the asymptotics $d\to\infty$ if
this limit is analytic for the two-particle functions. Since the neglected
contributions to the two-particle vertex are less singular than the leading
divergent $d\to\infty$ terms containing the diffusion and Cooper poles, we
can assume analyticity of the high-dimensional limit, even if we cannot
prove it rigorously. Using the high dimensional separation of singularities
of the full vertex into the irreducible ones we obtain from
Eq.~\eqref{eq:Phi-diffusion} at zero temperature for $q=0$
\begin{equation}\label{eq:eh-vertex-singular}
\Lambda^{RA}_{\mathbf{k} \mathbf{k}'}(E+\omega,E) =
\lambda^{RA}_{\mathbf{k} \mathbf{k}'}(E+\omega,E) +
\frac{2\pi n_F\lambda}{-i\omega +
D(\omega)(\mathbf{k} + \mathbf{k}')^2}
\end{equation}
where $\lambda$ is a measure of the disorder strength.  The function
$\lambda^{RA}_{\mathbf{k} \mathbf{k}'}$ is  nonsingular or at least less
singular than the Cooper pole singled out from the electron-hole
irreducible vertex, second term on the r.h.s. of
Eq.~\eqref{eq:eh-vertex-singular}. Notice that the asymptotic form of the
Copper pole in Eq.~\eqref{eq:eh-vertex-singular} is exact only if the the
Ward identity \eqref{eq:VWW-identity} is obeyed. The Cooper pole in
Eq.~\eqref{eq:eh-vertex-singular} is the generally anticipated singularity
of the electron-hole irreducible vertex \cite{Vollhardt80b}.

To test consistence of such singular behavior of the electron-hole
irreducible vertex we define a difference of two
retarded self-energies
\begin{subequations}
\begin{equation} \label{eq:SE_difference}
\Delta W_E(\omega) = \frac{1}N \sum_{\mathbf{k}}
[\Sigma^R(\mathbf{k},E-\omega) - \Sigma^R(\mathbf{k},E +\omega)]\ .
\end{equation}
We will be interested in its low-frequency behavior, i.~e., the asymptotic
limit $\lim_{\omega\to0}\Delta W_E(\omega)$. If the self-energy
$\Sigma^R(E)$ is an analytic function of the energy variable $E$, which
normally is the case within the energy bands, then $\Delta W_E(\omega)$
must be analytic at $\omega=0$ for almost all energies~$E$ for which
$\Im\Sigma^R(E)> 0$. If the self-energy is a continuous function of the
energy argument, then $\Delta W_E(0) = 0$. Finiteness of the derivative of
the self-energy demands $\lim_{\omega\to0}|\Delta W_E(\omega)/\omega|
<\infty$, etc.
 
We apply the Ward identity \eqref{eq:VWW-identity} to represent the r.h.s.
of Eq.~\eqref{eq:SE_difference} via the electron-hole irreducible vertex.
We obtain
\begin{multline}\label{eq:vertex-difference}
\Delta W_E(\omega) = \frac {-1}{N^2}\ \sum_{\mathbf{k}\mathbf{k}'}\left\{2i
\right. \\ \left. \times\left[\Lambda^{RA}_{\mathbf{k}
\mathbf{k}'}(E+\omega,E)  - \Lambda^{RA}_{\mathbf{k}
\mathbf{k}'}(E-\omega,E)\right] \Im G^R(\mathbf{k}',E)\right. \\ \left.  +
\Lambda^{RA}_{\mathbf{k}
\mathbf{k}'}(E+\omega,E)\left[G^R_{\mathbf{k}'}(E+\omega)  -
G^R_{\mathbf{k}'}(E)\right] \right. \\ \left. - \Lambda^{RA}_{\mathbf{k}
\mathbf{k}'}(E-\omega,E) \left[G^R_{\mathbf{k}'}(E-\omega) -
G^R_{\mathbf{k}'}(E)\right] \right\} \ .
\end{multline}
\end{subequations}
We use representation \eqref{eq:eh-vertex-singular} for the electron-hole
irreducible vertex  to derive the low-frequency asymptotics
$\lim_{\omega\to0}\Delta W_E(\omega)$. Actually we are interested only in
the most singular part being generated by the Cooper pole. It is a
straightforward task to perform momentum integration on a $d$-dimensional
hypercubic lattice leading to \cite{Janis03a}
\begin{multline}\label{eq:nonanalyticity}
\Delta W^{sing}_E(\omega)  \thickapprox K_d \lambda n_F^2 \\ \times
\begin{cases}
 \frac 1{\omega} \left|\frac{\omega}{D(\omega)k_F^2}\right|^{d/2}\  &
\text{for $d\neq4l$},\\[4pt] \frac 1{\omega}
\left|\frac{\omega}{D(\omega)k_F^2}\right|^{d/2} \ln\left|\frac
{D(\omega)k_F^2}\omega\right| & \text{for $d=4l$}.
\end{cases}
\end{multline}

In the diffusive phase, $D(\omega) = D>0$, that realizes only for $d>2$, we
find nonexistence of derivatives of the self-energy independently of the
energy position within the band. Realization of a self-energy with
nonexisting (diverging) $[(d-2)/2]$th derivative everywhere is
apparently unphysical for $d<\infty$.
 
The situation worsens if we evaluate $\lim_{\omega\to0}\Delta W_E(\omega)$
for the localized solution with the low-frequency dynamical diffusion
function from Eq.~\eqref{eq:diffusion-localized}. We then obtain
independently of the spatial dimension $\Delta W^{sing}_E(\omega) \sim
C_d/ \omega$. This is a catastrophe indicating nonexistence of the
retarded self-energy as a well defined mathematical function. This may not
happen. Hence, our picture based on the Ward identity
\eqref{eq:VWW-identity} resulting in the Cooper pole in the electron-hole
vertex, Eq.~\eqref{eq:eh-vertex-singular}, is inconsistent. The real
solution of the Anderson model must behave differently from the above
derived singular behavior.

The correct low-energy behavior of the electron-hole vertex can be assessed
from the exact asymptotic solution in high spatial dimensions. The limit
to high spatial dimensions leads to simplifications in momentum
convolutions so that one can solve the parquet equations for the
irreducible vertices $\Lambda^{eh}$ and $\Lambda^{ee}$ asymptotically
exactly \cite{Janis04a}. In this asymptotic solution the Ward
identity~\eqref{eq:VWW-identity} is not obeyed for nonzero real energy
differences $\omega\neq0$. A deviation from the Ward identity leads to a
modified diffusion (Cooper) pole. In the localized phase we then obtain
for $\omega,(\mathbf{k} + \mathbf{k}')^2\to0$ the electron-hole
irreducible vertex in the asymptotic form
\begin{equation}\label{eq:Phi_localized}
\Lambda^{RA}_{\mathbf{k}\mathbf{k}'}(E + \omega, E)\approx \frac
i{A(\omega)\omega} \ \frac {2\pi n_F\lambda} {1 + \xi^2(\mathbf{k} +
\mathbf{k}')^2} \
\end{equation}
 where $\xi^2 = iD(\omega)/A(\omega)\omega$ is the square of the
localization length. The limit $\lim_{\omega\to0}A(\omega)\omega$ becomes
finite inside the localized phase as well as the localization
length~$\xi$. We can see from Eq.~\eqref{eq:Phi_localized} that the
electron-hole irreducible vertex does not display a low-energy
singularity. It means that there is also no diffusion pole in the
electron-hole correlation function in the localized phase. This is
a necessary condition for the existence of the self-energy as a~well
behaved function along the real axis of the energy variable (Fermi
energy).

The fraction $|1/A(\omega)|\le 1$  generally determines the weight of the
diffusion pole. The static value $A(0)$ is finite in the diffusive phase,
increases with the disorder strength and diverges at the Anderson
localization transition. The renormalized density of states $n_F/A(0)$
represents the number of diffusive states, that is, extended states being
able to carry electron charge to long distances. These are the only states
contributing to charge transport, whereas $n_F$ is the density of all
available states at the Fermi energy $E_F$.
 
Nonconservation of the averaged particle density incurred by a violation of
the Ward identity~\eqref{eq:VWW-identity} in the high-dimensional solution
for the vertex function is a~consequence of the configurational averaging
\cite{Janis04b}. Validity of Eq.~\eqref{eq:VWW-identity} is conditioned by
completeness of Bloch waves for all relevant configurations of the random
potential. During the configurational  averaging   we lose the
spatially restricted localized states in the thermodynamic limit. To form a
representation Hilbert space we have to fix boundary conditions for the
eigenstates of the random Hamiltonian. The extended states vanish at the
volume boundary as $V^{-1/2}$, while the localized states vanish
exponentially $\exp\{-V/v_0\}$. The two types of states are orthogonal in
the thermodynamic limit. Restoring translational invariance restricts our
description only to states extended over the whole volume. The only
localized states surviving configurational averaging are the periodically
repeated ones. If there is a macroscopic portion of configurations with
spatially bounded states breaking translational invariance near the Fermi
energy, the number of available translational invariant states is less
than the total number of states. Hence, the probability to find the
electron in an extended state is less than one.

To conclude, we showed that the diffusion pole in the electron-hole
correlation function cannot survive beyond the mobility edge where all
states become localized. We used only exact and nonperturbative arguments
and found incompatibility of the Ward identity~\eqref{eq:VWW-identity},
implying a singular low-energy asymptotics of the electron-hole correlation
function, Eq.~\eqref{eq:Phi-diffusion}, with the existence of the
retarded self-energy as a well defined function of the energy variable. We
demonstrated that a necessary condition for the existence of a regular
(finite) self-energy in the localized phase is regularity of the
electron-hole irreducible vertex in the low-energy limit. Supported by the
asymptotic solution of the Anderson model in high spatial dimensions we
concluded that the weight of the diffusion pole in the metallic regime is
$1/A(0)\leq 1$. This weight decreases with increasing disorder strength
and vanishes at the localization transition. The most striking consequence
of our finding is that it is not the diffusion constant $D(0)$ but the
weight of the diffusion pole $1/A(0)\leq 1$ that is the relevant control
parameter for the Anderson localization transition.

Research on this problem was carried out within a project AVOZ1-010-914
of the Academy of Sciences of the Czech Republic and supported in part by
Grant No. 202/04/1055 of the Grant Agency of the Czech Republic.

\end{document}